\documentstyle[aas2pp4]{article}
\slugcomment{PASJ 1998, in press}
 
\lefthead{Y. Sofue}
\righthead{M82 Disk Truncation}
 
\begin{document}

\def\Msun{M_{\odot \hskip-5.2pt \bullet}}
\def\kms{km s$^{-1}$}

\title{IS M82 A DISK-TRUNCATED BULGE BY A CLOSE ENCOUNTER WITH  M81?}

\author{Yoshiaki Sofue\altaffilmark{1} }
\affil{1.Institute of Astronomy, University of Tokyo, Mitaka, Tokyo 181,
Japan}

%\and 

%\altaffiltext{1}{ }  

\begin{abstract}  

The rotation curve of the small-mass starburst galaxy M82 has  
a steep nuclear rise, peaking at 200 pc radius, 
and then, declines in a Keplerian fashion.
This rotation curve mimics that 
for a central bulge of spiral galaxies with high 
concentration of stellar mass. 
The declining rotation indicates that 
its extended disk mass is missing.
In order to explain this peculiar rotation characteristics,
we propose a hypothesis that M82 is a surviving central
bulge of a much larger disk galaxy, whose outer disk was truncated
during a close encounter with M81. 
We simulate a tidal truncation of the disk
of a companion galaxy by a tidal penetration through its more massive
parent galaxy. 
The model can well reproduce the observed peculiar
feature of M82. 

\end{abstract}


\section{Introduction}

CO observations of the starburst galaxy M82 with 
high-resolution and high sensitivity have indicated a 
prominent concentration of molecular gas in the central starburst region
as well as a gas disk toward the optical edges 
(\cite{nak87,lo87,loi90,sof92}).
Various peculiar properties associated with the high
concentration of molecular gas have been claimed to exist
(\cite{rie80,tel91,nak87,sof88}).
Kinematically, a declining rotation in the outer disk 
has been found obviously in this galaxy (\cite{you84}), 
and a Keplerian rotation has been suggested (\cite{sof92}).
M82 has also an extended HI tail, indicating a strong interaction
with the giant spiral M81 (\cite{cot77,got77,app81,bro91}).
Moreover, high-resolution VLA observations in the HI line emission 
have provided an accurate position-velocity diagram in the outer
disk (\cite{yun93}). 

In this paper we integrate these CO and HI position-velocity data
to obtain an accurate rotation curve of M82, and discuss
the Keplerian rotation, which is very peculiar and exceptional
for a disk galaxy. 
Rotation curves are the principal tool to discuss the
kinematics and mass of galaxies, and have been investigated extensively 
based on optical (H$\alpha$) and  HI-line data for outer disks
(\cite{rub82,per96}),
and on CO-line data for nuclear regions (\cite{sof96,sof97}).
However, no galaxy, except M82, has been known to exhibit 
a Keplerian-like declining rotation within a few kpc radius.   
In order to explain this peculiar kinematics, we  
simulate the tidal interaction between M82 and M81,
and propose a tidal truncation model of M82's outer disk.
The distance to M82 is assumed to be 3.25 Mpc (\cite{tam68}).

\section{CO and HI-Line Rotation Curves}
 
Fig. 1 shows a superposition of the highest-resolution 
position-velocity diagrams
in the CO ($J=2-1$) (\cite{sof92}) and HI-line emission (\cite{yun93}). 
It is remarkable that the HI and CO emission regions are well mixed. 
Note that the brightest region in CO in the central 0.5 kpc is 
observed in absorption in the HI line.
Such a mixed existence of HI and CO is quite unique, different from 
the common interstellar property in other galaxies, where HI and CO 
gases are more clearly separated by a 
molecular front (\cite{sof94,hon95}).

\placefigure{fig1}

The most prominent kinematical feature observed in Fig. 1 
is the declining rotation velocity.
In order to obtaine the rotation curve,
we apply the envelope-tracing method (\cite{sof96,sof97})
to this diagram, and the result is shown in Fig. 2 by the thick line.
The dotted thin line in Fig. 2 shows an HI rotation curve,
as obtained from an intensity-averaged velocity field of HI gas (\cite{yun93}),
which indicates slightly slower values than those from the
envelope-tracing.
Note also that the HI data beyond 4$'$ are only for the north-eastern side.
Beyond 5$'$ radius, the galaxy has no clear disk component, and it is
difficult to trace any sysetematic velocity variation as a function of
the radius, being merged by the distorted HI envelope (\cite{cot77,got77,yun93}).

The rotation velocity of M82 increases steeply near the the center, 
and attains a sharp maximum at 200 pc radius. 
The rotation velocity, then, declines monotonically to a velocity
as low as 50 \kms\ at 4 kpc radius.
It is not likely that this behavior is due to a warping of a disk
with a flat rotation, because the galaxy is almost edge-on, 
and there is no sign of any strong warping to become face-on within 
a few kpc radius. 
We superpose a Keplerian rotation curve by a thin line in Fig. 2, 
which corresponds to a point mass of $4.6\times10^9\Msun$. 
The dotted line indicates a rotation curve of a sphere whose
surface density decreases exponentially with a scale
radius of 0.5 kpc, which fits the observation fairly well.
Thus, the rotation characteristics of M82 is quite different from those
for other spiral galaxies, which are generally flat from the central
few kpc to the outer disk (\cite{rub82,bos81,per95,per96,sof96,sof97}).
 
\placefigure{fig2}

Fig. 3 compares the rotation curve of M82 with those for the
nearby Sb spiral M81 (\cite{sof97}) and of our Galaxy (\cite{hon97}), 
both of typical massive Sb type.
In Fig. 3 we also compare it with those obtained for
37 nearby galaxies by thin lines (\cite{sof97}). 
The rotation curve of M82
is indeed exceptional: It has no flat part at all.
It is more peculiar, if it is compared with those of
dwarf galaxies with similar masses such as NGC 4631, showing a
gradually increasing rotation (e.g., \cite{per96}).
We emphasize, however, that the steep nuclear rise and sharp maximum
of the  rotation curve at 200 pc radius resembles that for a 
central bulge of a massive galaxy like M81 or the Milky Way, whose 
centrally peaking component can be decomposed into a massive bulge.

\placefigure{fig3} 

\section{Tidal Encounter and Truncation of a Disk}

In our earlier paper (\cite{sof92}), we have suggested
that the Keplerian behavior of M82's rotation might be
due to a truncation of an originally existed disk
during a close encounter with M81.
In order for the outer disk beyond 2 kpc to be truncated, 
we may assume that the tidal force of M81 
was strong enough even  at a radius of $r=r_t\sim2$ kpc. 
Then the perigalactic distance between M82 and M81, $R$, can be related
to their mass and $r_t$ through
$(r_t/R)^3 \sim M_{82}/M_{81}$, where $M_{82}$ and $M_{81}$ 
are the masses of the two galaxies. 
For $r_t \sim 2$ kpc,  $M_{82} \sim 5 \times 10^{9} \Msun$ and 
$M_{81} \sim  10^{11}\Msun$, we obtain $R\sim 4$  kpc. 
 
In order to investigate how a tidal disruption of outer disk has occurred, 
we  perform a simple simulation of the tidal encounter. 
We assume that a larger galaxy (M81) has a fixed potential similar
to that of the Milky Way, which comprises a bulge, disk and massive 
halo  (Table 1).
The tidally disrupted galaxy (M82) is assumed to be represented by  
single a Plummer's potential.
The rotation axes of the two galaxies are taken to be parallel to each other,
and are also parallel to the orbital angular momentum: 
Namely, the encounter occurs in the same galactic plane of the 
two galaxies in a direct sense.

The smaller galaxy's disk is represented by test particles, 
which are initially distributed in a rotating ring with a radius equal to 
the characteristic radius of the Plummer potential.
Here, the potential is written in the form of
$$
\Phi=GM \lbrack r^2 + a(\sigma_r)^2 \rbrack ^{-1/2},
$$ 
where $M$ is the mass of the disturbed galaxy, 
and the Plummer radius $a(\sigma_r)$ is proportional to the 
mean radius of the test particle distribution, defined by
$$
a(\sigma_r)=a^0 ~ {\sigma_r \over \sigma_r^0}.
$$
The mean radius $\sigma_r$ of test particles around the center of 
M82 is given by
$$
\sigma_r = {{\Sigma \Delta r_j} \over N},
$$
where $\Delta r_j$ is the spatial displacement of $j$-th particle
from the center of M82.
The total mass is taken to be proportional to $a^0$, so that the
initial peak velocity in the galaxy becomes constant (180 \kms) 
for any value of $a^0$.
This modification of the Plummer's potential represents a
semi-selfgravitating deformation of the disk.
If the galactic disk is tidally distorted, 
the particle distribution becomes more dispersed, 
resulting in an increase of the disk radius,
and therefore, in a decrease of the centrifugal force within the disk.  
The present test-particle simulation might be too simple to
be compared with the $N$-body self-gravitating simulations of
galaxy-galaxy interaction. 
However, the neglection of 
selfgravity will be not important in the present
calculation, which aims at estimating the tidal disruption radius, 
because the tidal impact occurs in a time much 
shorter than the self-gravity regulating time scale, which is about
several rotation periods.

We present an example of the calculations in Fig. 4.
The smaller galaxy (M82) is put at 200 kpc away from the center
of M81 with an impact parameter of 50 kpc and a velocity of 30 km/s.
M81 orbits around M82 with a peri- and apo-galactic distances of 3 and
220 kpc, respectively, in a period of 4 Gyr.
If the Plummer radius is larger than 1.2 kpc, the disk of the smaller galaxy 
is almost totally disrupted during the close encounter, as is represented by 
a sudden increase of the Plummer radius within $\sim 0.1$ Gyr.
On the other hand, if the Plummer radius is smaller than 1 kpc.
the bulge is only disturbed, but the Plummer radius remains almost unchanged.
In actual bulge, however, stars' orbits would be more radial, 
so that we also examined counter-rotating cases, but obtained
almost the same result.
We also calculated the same for a fixed Plummer radius, and 
obtained almost the same behavior of the radius of test particle
distribution as above.
By this simple simulation, we have 
shown that M82's disk component is easily
truncated during a close encounter, while the bulge 
can survive the tidal disruption.

\placefigure{fig4}

We have analyzed and modeled the data on the assumption of 
circular rotation for the  first-order approximation.
We have no mean to analyze a higher-order motion,
because the galaxy is edge-on.
However, it is possible that the tidal interaction
has caused a bar and non-circular motion (\cite{nog88,tel91,yun93}). 
If such a bar is looked at side-on, the position-velocity diagram would
behave more rigid-body like, manifesting the bar's pattern  speed
(e.g. \cite{wad94}). 
This is, however, not the case in the observed data.
If the bar is looked at end-on, although this chance is small,
a steeper rise  and fall of radial velocity than that expected from
a circular motion  would be superposed.
In fact, if we look at the the rotation curve in Fig. 2 more carefully,
it appears to decrease a little more rapidly than Keplerian, which
might be due to a bar.
Bar related kinematics is beyond the scope of this paper.
However, it would not affect the result of this paper, which discuss
the lack of an outer flat-rotation disk.

\section{ Discussion}

Evidences for tidal interaction between M82 and M81 have been obtained.
The two galaxies, together with NGC 3077 and NGC 2976, are embedded in
a huge HI-gas envelope.
The two galaxies are linked by an HI bridge, and 
a tidal tail extends from M81 toward the south  
(\cite{got77,app81,yun93}).
The molecular gas disk is also largely extended and distorted, 
extending toward the halo of M82 as high as $\sim 1$ kpc, and to the south 
(\cite{olo84,you84,nak87,lo87,sof92}).
Moreover, the amount of molecular gas is extremely large, 
anomalous for the small total mass of M82, and
is thought to be the cause of its intense starburst 
(\cite{rie80,tel91,sof88}).
On the other hand, M81 is known for its gas depression in the center
 (\cite{vdh79,sag91}). 

We summarize a possible scenario for the M82-M81 interaction as follows.
M82 was a larger and more massive spiral galaxy, 
containing HI gas in the outer disk and dense
molecular gas in the central region.
Through the close encounter with M81, when M82 penetrated the disk of M81, 
the outer disk of M82 was tidally truncated, but the bulge 
and nuclear disk have survived the tidal disruption.
The truncated disk may have become the HI envelope and tails 
in the M81-M82 system.
The central gas disk of M82 was dense enough, and ram-stripped the 
gas disk of M81 away, which was accumulated in the center of M82. 
This has caused the high-density molecular disk in M82, and a starburst. 
In contrast, M81 evolved into such a galaxy that contain little gas
in the center.

\acknowledgments
 
%\clearpage

%\begin{table*}
\begin{table}
\begin{center}
\tablenum{1}
\caption{Potential parameters for M81 and M82  \label{tbl-1}} 
\begin{tabular}{crrr}\\
\tableline 
Galaxy &Component 	&Mass $M_i$ 	& Radius $a_i^0$ 	 \\ 	
	& 	   	&($10^{11}M\odot$)&(kpc) 		 \\
\tableline 
M81	&Bulge 		&0.1 		&0.5		\\
 	&Disk 		&1.1		&$7\times0.5$\\ 
 	&Halo 		&1.3		&20		\\
	&Total		&2.5		&		\\
M82	&Bulge		&0.1		&0.4		\\
	&Disk		&1.0		&5.0		\\
	&Total		&1.1		&	\\
\tableline
\end{tabular}
\end{center}
%\end{table*}
\end{table}

%\clearpage 

%\clearpage

\figcaption[fig1.eps]
{Position-velocity diagram along the major 
axis of M82 in CO (J=2-1) (thin contours), 
superposed on that in HI (hatched).
\label{fig1}}

\figcaption[fig2.eps]
{Rotation curve of M82. Rotation velocities for a point mass
(thin line) and an exponential-surface density 
sphere (dashed line) are indicated. 
\label{fig2}}

\figcaption[fig3.eps]
{Rotation curve of M82 (full line) 
compared with those of M81 (dashed line) and the Milky Way (dotted line),
as well as with those for other spiral galaxies 
(thin lines). 
\label{fig3}}

\figcaption[fig4.eps]
{Plot of the Plummer radius  during a close encounter as a function of time.  
The galaxy is disrupted when the Plummer radius is larger than 1 kpc.
\label{fig4}}


\begin{thebibliography}{}
\def\r{\bibitem} 
\baselineskip 0pt 
 
\r[Appleton et al 1981]{app81} Appleton, P.N., Davies, R.D., 
and Stephenson, R.J. 1981, \mnras, 195, 327
\r[Bosma 1981]{bos81} Bosma, A. 1981, \aj,  86, 1825.
\r[Brouilet et al 1991]{bro91} Brouilet, N., Baudry, A., 
Combes, F., Kaufman, M., and Bash, F. 1991, \aap, 242, 35.
\r[Clemens 1985]{cle85} Clemens, D. P. 1985, \apj, 295, 422 
\r[Cottrell 1977]{cot77} Cottrell, G. A. 1977, \mnras, 178, 577.
\r[Gottesman and Weliachew 1977]{got77} Gottesman, S. T., and Weliachew, L. 
1977, \apj,  211, 47.
\r[Honma and Sofue 1997]{hon97} Honma, M., and Sofue, Y., 1997, PASJ, in press.
\r[Honma et al 1995]{hon95} Honma, M., Sofue, Y., and Arimoto, N. 1995, \aap, 304, 1.
\r[Lo et al 1987]{lo87} Lo, K.Y., Cheung, K.W., Masson, C.R., 
Phillips, T.G., Scott, S.L., Woody, D.P.: 1987, \apj, 312, 574
\r[Loiseau et al 1990]{loi90} Loiseau, N., Nakai, N.,  Sofue, Y., 
Wielebinski, R., and Klein, U. 1990,  \aap,  228, 331. 
\r[Nakai et al 1987]{nak87} Nakai, N., Hayashi, M., Handa, T., 
Sofue, Y., and Hasegawa, T., 1987, \pasj, 39, 685
\r[Noguchi 1988]{nog88} Noguchi, M. 1988, \aap, 203, 259.
\r[Olofson and Rydbeck 1984]{olo84} Olofson, H., and Rydbeck, G. 
1984, \apj, 247, 473.
\r[Persic and Salucci 1995]{per95} Persic, M., and Salucci, P.  1995, \apjs,
\r[Persic et al 1996]{per96} Persic, M., Salucci, P., Stel, F.  1996, \mnras, 281,  27 
99, 501. 
\r[Rieke et al 1980]{rie80} Rieke, G.H., Thompson, R.I., Low, F.J., 
and Tokunaga, A.T. 1980, \apj, 238, 24
\r[Rubin et al 1982]{rub82} Rubin, V. C., Ford, W. K., 
Thonnard, N. 1982, \apj, 261, 439
\r[Sage and Westpfahl 1991]{sag91} Sage  L. I.,    Westpfahl  D. J. 1991, \aap, 242, 371.
\r[Sofue 1988]{sof88} Sofue, Y.: 1988, in  Galactic and Extragalactic Star 
Formation, eds. R.E. Pudritz and M. Fich (NATO ASI Series), Reidel, Dordrecht, p. 409
\r[Sofue 1996]{sof96} Sofue, Y. 1996, \apj, 458, 120
\r[Sofue 1997]{sof97} Sofue, Y. 1997, PASJ, 49 
\r[Sofue et al 1992]{sof92} Sofue, Y., Reuter, H.-P., Krause, M., Wielebinski, R., 
and Nakai, N., 1992, \apj, 395, 126.
\r[Sofue et al 1994]{sof94} Sofue, Y., Honma, M., Arimoto, N. 1994, \aap, 296, 33.
\r[Tamman and Sandage 1968]{tam68} Tammann, G.A., and Sandage, A.R.: 1968, \apj,  151,825.
%\r[Telesco and Harper 1980]{tel80} Telesco, C.M. and Harper, D.A., 1980, \apj, 235, 392.@
\r[Telesco, et al 1991]{tel91} Telesco, C.M. , Campins, H., Joy, M., Dietz,
K., and Decher, R. 1991, \apj, 369, 135.
\r[van der Hulst 1979]{vdh79} van der Hulst, J. M., 1979, \aap, 75, 97.
\r[Wada et al 1994]{wad94} Wada, K., Taniguchi, Y., Habe, A., and Hasegawa, T., 1994, \apj, 437, L123.
\r[Young and Scoville 1984]{you84} Young, J.S., Scoville, N.Z.: 1984, \apj, 187, 153
\r[Yun et al 1993]{yun93} Yun  M. S.,  Ho  P.T.P, and   Lo  K.Y. 1993, ApJ 411,  L17


\end{thebibliography}
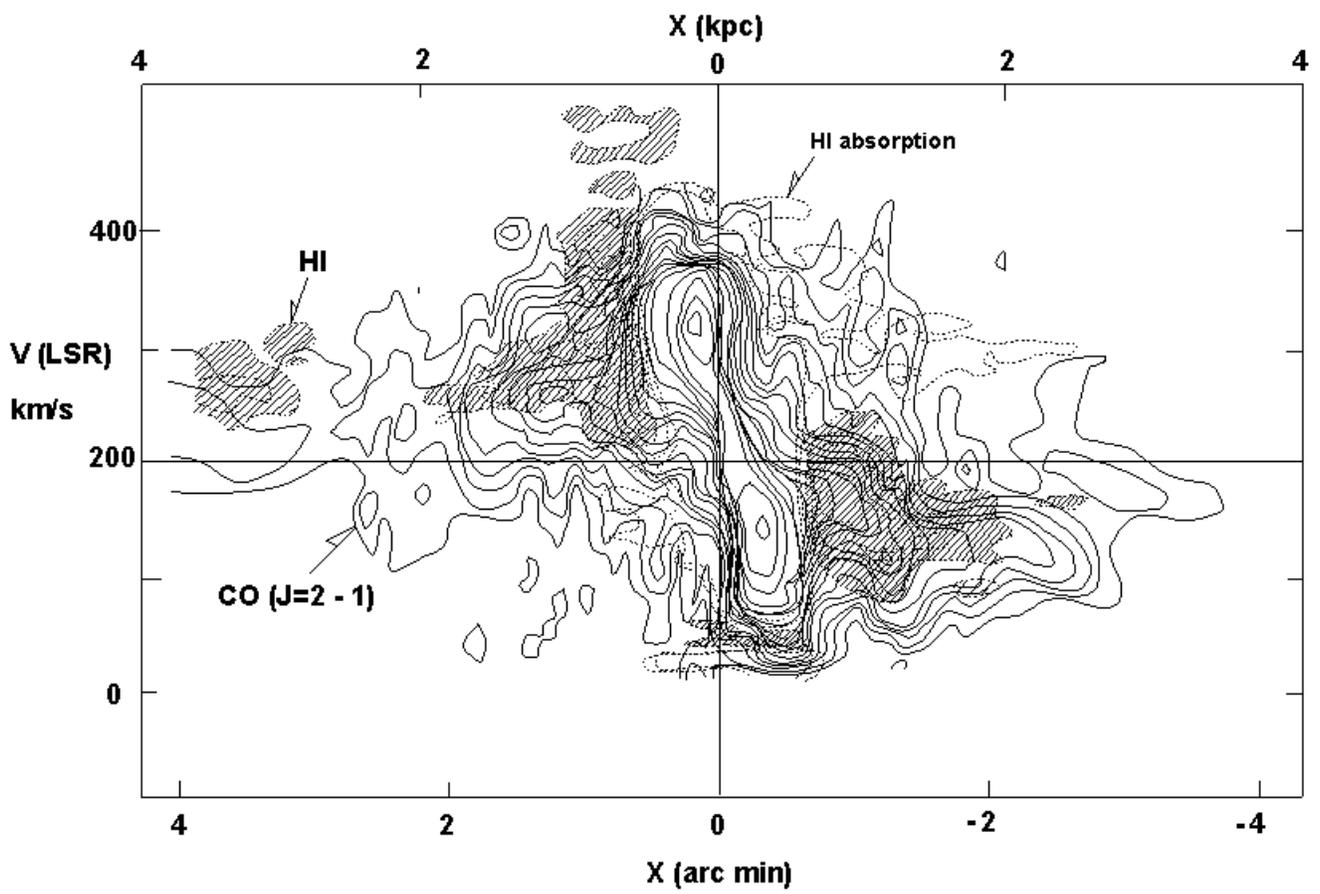
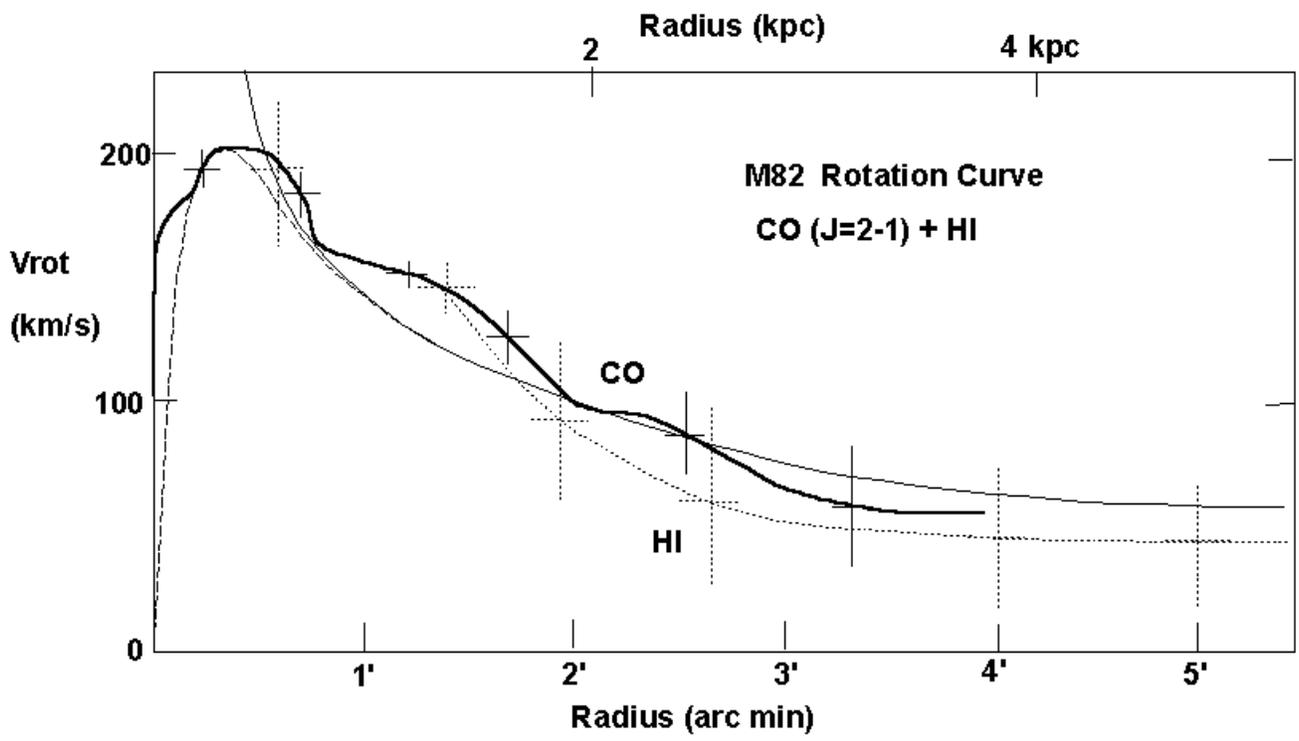
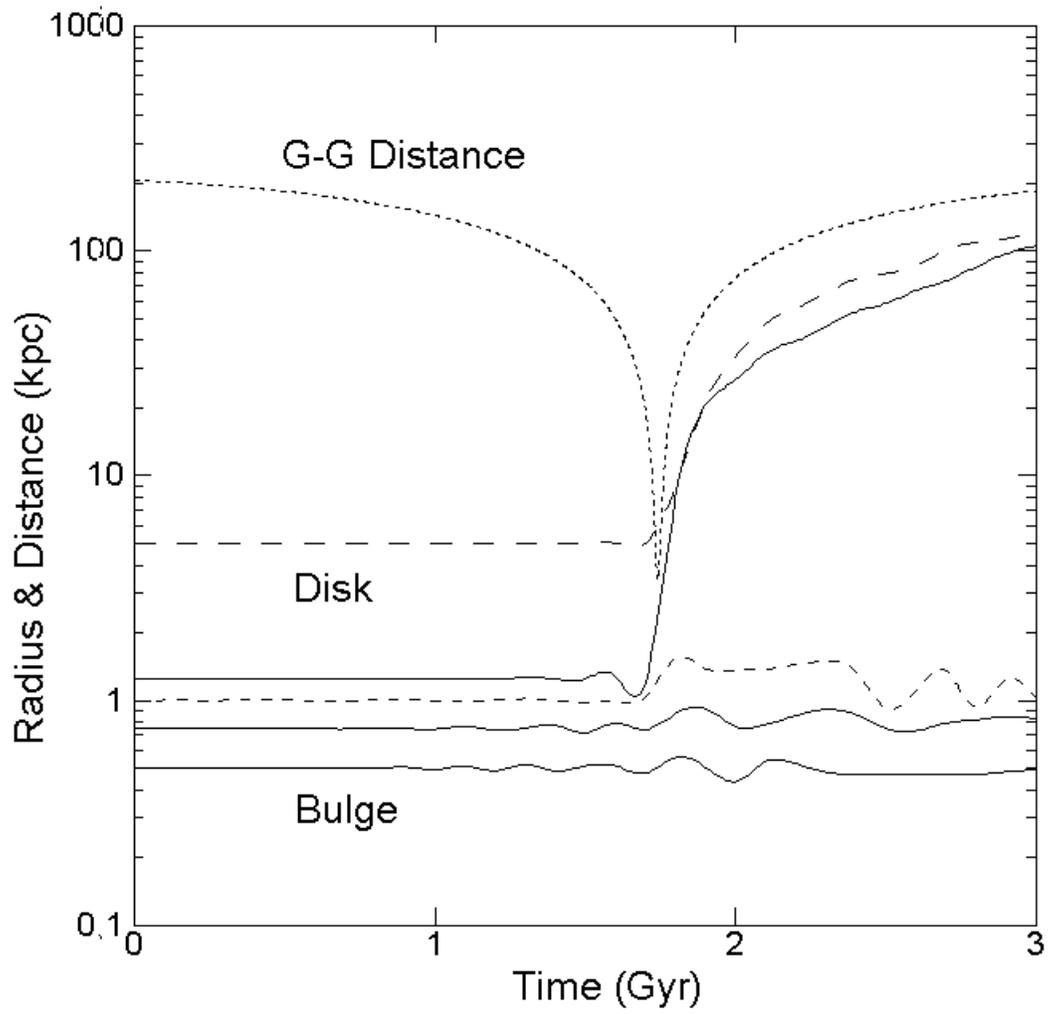
\end{document}